\documentclass[pra,aps]{revtex4}

\usepackage{german,epsfig,pstricks}
\usepackage{graphicx}
\usepackage{amsmath,amssymb}

\newcommand{\tr}{\operatorname{tr}}
\newcommand{\diag}{\operatorname{diag}}
\newcommand{\ket}[1]{\left|#1\right\rangle}      

\def\trace{{\rm tr}\;}
\newcommand{\Matrix}[2]{\left( \begin{array}{#1} #2 \end{array}
  \right)}

\newcommand{\beq}{\begin{equation}}
\newcommand{\beqa}{\begin{eqnarray}}
\newcommand{\eeq}{\end{equation}}
\newcommand{\eeqa}{\end{eqnarray}}
\newcommand{\nbeqa}{\begin{eqnarray*}}
\newcommand{\neeqa}{\end{eqnarray*}}
\def\bal{\begin{align}}
\def\eal{\end{align}}

\def\e{{\rm e}}

\def\CC{{\rm\kern.24em \vrule width.04em height1.46ex depth-.07ex
\kern-.30em C}}
\def\P{{\rm I\kern-.25em P}}

\def\CC{{\rm\kern.24em \vrule width.04em height1.46ex depth-.07ex
\kern-.30em C}}
\def\PP{{\rm I\kern-.25em P}}
\def\RR{{\rm
         \vrule width.04em height1.58ex depth-.0ex
         \kern-.04em R}}
\def\id{{\rm 1\kern-.22em l}}
\def\ZZ{{\sf Z\kern-.44em Z}}
\def\NN{{\rm I\kern-.20em N}}

\def\G{{\cal G}}

\def\P{{\cal P}}

\def\T{{\cal T}}

\begin{document}

\date{\today}

\title{Integrable spin-boson models descending from rational six-vertex models}

\author{L. Amico}
\affiliation{MATIS-INFM $\&$ Dipartimento di Metodologie Fisiche e Chimiche
  (DMFCI), Universit\'a di Catania, viale A. Doria 6, I-95125 Catania, Italy}
\author{H. Frahm, A. Osterloh}
\affiliation{Institut f\"ur Theoretische Physik, Leibniz Universit\"at
  Hannover, Appelstra\3e 2, D-30167 Hannover, Germany}
\author{and G.A.P. Ribeiro}
\affiliation{Theoretische Physik, Universit\"at Wuppertal,
Gau\3stra\3e 20, D-42097 Wuppertal, Germany $\&$ \\ Universidade
Federal de S\~ao Carlos, Departamento de Fisica, CP
  676, 13565-905 S\~ao Carlos-SP, Brazil}

\begin{abstract}
We construct commuting transfer matrices for models describing the 
interaction between a single quantum spin and a single bosonic mode using 
the quantum inverse scattering framework.  The transfer matrices are 
obtained from certain inhomogeneous rational vertex models combining 
bosonic and spin representations of $SU(2)$, subject to non-diagonal 
toroidal and open boundary conditions.  Only open boundary conditions are 
found to lead to integrable Hamiltonians combining both rotating and 
counter-rotating terms in the interaction.  If the boundary 
matrices can be brought to triangular form simultaneously, the 
spectrum of the model can be obtained by means of the algebraic Bethe 
ansatz after a suitable gauge transformation; the corresponding  
Hamiltonians are found to be non-hermitian.  Alternatively, a 
certain quasi-classical limit of the transfer matrix is considered where 
hermitian Hamiltonians are obtained as members of a family of commuting 
operators; their diagonalization, however, remains an unsolved problem.
\end{abstract}

\maketitle

\section{Introduction}

Models describing the interaction between bosonic modes and spin degrees of freedom play an important role in physics.
The prototype application is provided by atom-radiation 
interactions~\cite{COHEN},
where the atomic dipole is described by an effective
spin interacting  with a single mode electric field\cite{JAYNES,TAVIS}.
More recently, this class of
models has been considered to investigate  various aspects in
mesoscopic physics where a given two-level system interacts with a 
bosonic environment\cite{WEISS}.
Furthermore, we mention applications to analyze the decoherence in
superconducting circuits\cite{DEVORET,FALCI}, and to describe   
ions in harmonic traps~\cite{ION-TRAPS}. 
Both these quantum devices are of
potential importance for quantum computation\cite{MOOIJ,RECENTEXP}. 
Finally, applications have emerged in quasi-2D semiconductors in 
transverse magnetic field~\cite{DATTA,SCHLIEMANN,SPIN-HALL}.

In this paper we focus on models describing a
single spin $S$ interacting with a single mode boson field
which were  introduced originally by Jaynes, Tavis and 
Cummings~\cite{JAYNES,TAVIS}.
In two cases the corresponding model Hamiltonian 
is known to be exactly diagonalizable with elementary means:
for $S\rightarrow \infty$ where the model is quadratic~\cite{EMARY}
and 
for weak or resonant interactions, where the Rotating Wave Approximation (RWA) 
can be applied~\cite{JAYNES,TAVIS,LIEB}. 
In the traditional framework of laser physics, 
the RWA assumes that the relevant dynamics is  given by coherent 
oscillations of the population of the atomic energy levels. 
Therefore only photon emissions
accompanied by an excitation of the atom (and vice versa), described by
operators of the type $a^\dagger S^-$, are taken into account 
in the atomic dipole coupling  with  electric 
field $\vec{E}\cdot \vec{d}\sim  (a+a^\dagger) S^x$. The so-called  
counter-rotating terms $a^\dagger S^+ + a S^-$,  
{\it although contained in the coupling}, 
can be neglected  at resonant cavity frequency $\Omega_0$ since 
they induce a short time dynamics 
(compared to the rotating terms $a S^+ + a^\dagger S^-$); as a result of the approximation the operator $S^z+a^\dagger a$ is conserved. Such a picture 
is robust for weak fields and as long as the cavity mode can be tuned 
conveniently. In the most recent applications, however, one or even both  
these conditions are not met. In superconducting circuits, for example,  
the effective spin-boson coupling to the electromagnetic field of the device 
can be naturally large, thereby involving multi-boson processes, 
which are characteristic for the presence of counter 
rotating terms\cite{MULTI-BOSON}; 
furthermore numerical inspections indicate that such terms can bend the  
energy levels whose slope should be controlled reliably  
to study how the noise affects the coherence in the system~\cite{MAUGERI}. 
Finally we mention that in semiconductors, 
rotating and counter rotating terms result from Rashba and 
Dresselhaus spin-orbit interactions; the physical origin 
of the spin transport in such systems is an open problem that is 
intensively studied~\cite{SPIN-HALL}.  

For finite spin representations and generic spin-boson interaction the
Hamiltonian is a non trivial matrix of infinite rank.
The main complication comes from the absence of a conserved ``number operator''
of the type $S^z+a^\dagger a$ as in the RWA
that would have made the problem finite dimensional 
-- $a^\dagger a$ is the bosonic number operator. 
In this sense, using the language of spin chains,
the problem will have the complexity of an XYZ chain. 
We will speak of a spin-boson model to contain rotating terms only, if the 
Hamiltonian action only connects a subspace of finite dimension.
It is worth noting that
it is the {\it simultaneous presence} of so called rotating, $S^\dagger a$, 
and counter-rotating terms $S^\dagger a^\dagger$ 
that makes the Hamiltonian non-trivial in the sense that a reduction to
a finite dimensional subspace is impossible.

In this work we study integrable spin-boson models that can be 
constructed by means of
the quantum inverse scattering method (QISM) \cite{KOREPIN-BOOK}.
The central object is the transfer matrix, that it is a generating
function of integrals of the motion, including the Hamiltonian by definition.
The formalism includes a certain freedom to choose (twisted) periodic
boundary conditions and open boundary conditions, where two boundary
matrices can be included, representing boundary magnetic fields in the
standard formalism for spin chains~\cite{SKLYANIN,VE}.
Twisted boundary conditions have been investigated for XXX-symmetry,
i.e. for models emerging from the XXX $R$-matrix in \cite{AH}
and hermitian spin boson models have been created that still have
a conserved number operator, indicating that the apparent counter-rotating
terms are spurious.
In the present work, open boundary conditions for XXX-type symmetry are
analyzed, where two non-parallel boundary matrices are meant to break
the XXX symmetry completely. Integrable open boundaries,
corresponding to dynamic boundary matrices (i.e. depending on the
spectral parameter) obey the reflection equation due 
to Sklyanin~\cite{SKLYANIN}.
Two approaches are used.
First, a bosonic Lax matrix, relative to the XXX $R$-matrix is
employed and the QISM and algebraic Bethe ansatz is elaborated for this case.
Second, the spin-boson Hamiltonian is obtained from
a two-site spin chain with mixed representations by virtue
of a 'large S limit' (algebraic contraction) of one of the spins to a bosonic degree of freedom.

The paper is laid out as follows. In the next section we summarize the ingredients of the reflection
equations to deal with integrable models with generic boundaries. In the 
Section \ref{spinboson} we present
the spin boson transfer matrix. The diagonalization will be pursued by 
a 'direct' algebraic Bethe ansatz
(Section \ref{ABA-direct}) and by means of the algebraic contraction of
a suitable auxiliary spin 
problem (Section \ref{ABA-contraction}). The Hamiltonian models extracted from the transfer
matrix  will be presented in the Section \ref{hamiltonians}. The Section \ref{conclusion} is devoted to the
conclusions.

\section{Integrable boundary conditions}
\label{boundary-conditions}

In this section, we review the basic aspects of the quantum inverse scattering
method providing integrable theories with  non-trivial boundary conditions.

The central object of the method is the transfer
matrix $t(\lambda)$, which can be conveniently written as the trace over an
auxiliary space ${\cal A}$ of an ordered product of $L$ distinct Lax matrices
${\cal L}_{j}(\lambda)$ acting in the site $j$. It can be explicitly written as
\begin{equation}
t(\lambda)=\tr_{{\cal A}}[{ T(\lambda)}]
\label{transferT1},
\quad
T_{}(\lambda)={\cal L}_{L}(\lambda){\cal L}_{L-1}(\lambda)
\dots {\cal L}_{1}(\lambda),
\end{equation}
where $T(\lambda)$ is called monodromy matrix.

The $R$-matrix and Lax matrices ${\cal L}_{j}(\lambda)$ are solutions of
the Yang-Baxter equations
\begin{eqnarray}
R_{12}(\lambda-\lambda') {\cal L}_j^{(1)}(\lambda) {\cal L}_{j}^{(2)}(\lambda') &=&
{\cal L}_{j}^{(2)}(\lambda')  {\cal L}_{j}^{(1)}(\lambda) R_{12}(\lambda- \lambda'),
\label{fundrel} \\
R_{12}(\lambda) R_{13}(\lambda+\lambda') R_{23}(\lambda') &=& R_{23}(\lambda') R_{13}(\lambda+\lambda') R_{12}(\lambda)\; ,\label{yangbaxter}
\end{eqnarray}
where the suffix $m$ in ${\cal L}_j^{(m)}$ indicates that in a multiple tensor
product of auxiliary spaces, the Lax matrix acts non-trivially only on
the $m$-th copy.
A rational solution of (\ref{yangbaxter}) is the following $R$ matrix
\begin{equation}
R(\lambda)=\left(
\begin{array}{cccc}
\lambda + \eta & 0 & 0 & 0 \\
 0 & \lambda & \eta & 0 \\
 0 & \eta & \lambda & 0 \\
 0 & 0 & 0 & \lambda + \eta
\end{array}
\right),
\label{rmatrix}
\end{equation}
where $\eta$ is a $\CC-number $ named quasi-classical parameter. The Yang-Baxter equation for the Lax matrices (\ref{fundrel}) provides the condition to the commutation property of the transfer matrix (\ref{transferT1}) and give rise to the fundamental commutation relations (FCR) for their operator entries.

In the realm of the QISM boundary terms can be considered. The
simplest way to do it is to construct the monodromy matrix in
terms of modified Lax matrices ${\cal L}(\lambda)\rightarrow {\cal
G} {\cal L}(\lambda)$, ${\cal G}$ being a matrix of constants
(with respect to the spectral parameter)\cite{VE}. The
corresponding theory is still integrable so far the quadratic
relation (\ref{fundrel}) is satisfied. For that it suffices that
\begin{equation}
[R_{12}(\lambda), {\cal G}^{(1)} {\cal G}^{(2)}]=0 \;
\label{Gcondition}
\end{equation}
For the case of $R$-matrix (\ref{rmatrix}) this relation is
satisfied by any $2\times 2$ matrix with entries $g_{11}, g_{12},
g_{21}, g_{22}$ independent of the spectral parameter.

It turns out that such modified monodromy matrices correspond to
theories with  toroidal boundary conditions. In the case where the
boundary matrix ${\cal G}$ is diagonal the corresponding transfer
matrix (\ref{transferT1}) can be diagonalized with very little
difference from the periodic case because it does not change in a
drastic way the properties of the monodromy matrix elements. In
general this does not occur when ${\cal G}$ is non-diagonal.
Nevertheless, for the rational solutions of the Yang-Baxter
equation (\ref{yangbaxter}), it is possible to apply a combination
of suitable auxiliary and quantum space transformations which
relate the non-diagonal boundary problem with a diagonal one
\cite{RM}. These transformations are the irreducible
representations of the symmetry of the auxiliary and quantum space
of the $\cal L$-operator ${\cal L}_{j}(\lambda)$. Such invariance
of the Lax matrix can be expressed by \cite{HO}
\begin{equation}
{\cal{L}}_{j}^{(1)}(\lambda)= \left[ {\cal G}^{(1)} U_j
\right]^{-1} {\cal{L}}_{j}^{(1)}(\lambda) \left[ {\cal G}^{(1)}
U_j \right], \label{invariance}
\end{equation}
where $U_j$ stand for the irreducible representation of the
symmetry of the quantum space \cite{RM}. For instance, if the
quantum space is invariant by spin-$S$ representation of $SU(2)$,
the relation (\ref{invariance}) is satisfied by
\begin{equation}
U_j= \left(\begin{array}{cccc}
                        u_{S,S} & u_{S,S-1} & \cdots & u_{S,-S} \\
                        u_{S-1,S} & u_{S-1,S-1} & \cdots & u_{S-1,-S} \\
                        \vdots & \vdots & \ddots & \vdots \\
                        u_{-S,S} & u_{-S,S-1} & \cdots & u_{-S,-S} \\
                        \end{array}\right),
\label{quantumtransf}
\end{equation}
where some of the matrix elements $u_{i,j}$ are
\begin{equation}
u_{S,S-k} = g_{1,1}^{2S_j-k}g_{1,2}^{k}
\sqrt{\frac{(2S)!}{k!(2S-k)!}},
\end{equation}
while the remaining satisfy the following recurrence relation
\begin{eqnarray}
u_{S-k-1,S-n}=\frac{(g_{1,1} g_{2,2}- g_{1,2}
g_{2,1})}{g_{1,1}^{2} }
\sqrt{ \frac{n(2S-n+1)}{(k+1)(2S-k)}} u_{S-k,S-n+1} \nonumber \\
+\left( \frac{g_{2,1}}{g_{1,1}} \right)^{2}
\sqrt{\frac{k(2S-k+1)}{(k+1)(2S-k)}}u_{S-k+1,S-n}
+\frac{g_{2,1}}{g_{1,1}} \frac{2(S-k)}{\sqrt{(k+1)(2S-k)}}
u_{S-k,S-n},
\end{eqnarray}
and $k,n$ are integers satisfying $ k,n=0, \dots,2S$.

As the transfer matrix (\ref{transferT1}) gives origin to a commutative
family of operators, it can be regarded as generating functional of the
integrals of the motion of the quantum theory, among which one of them can be
taken as the Hamiltonian. For example the logarithmic derivative of
$t(\lambda)$ provided a protocol to demonstrate the integrability of
a variety of local models (that is with nearest neighbor interactions)\cite{KOREPIN-BOOK}. For later use we remark that  an arbitrary function of $t(\lambda)$
or e.g. its coefficients in a Taylor series in $\lambda$
can be taken to be the Hamiltonian.

A different class of lattice models can be obtained focusing on the
quasi-classical parameter $\eta$. Specifically, integrable long-range
Hamiltonians, as the Gaudin and the BCS model, emerge from the so called
quasi classical limit $\eta\to 0$ of
vertex models~\cite{SKLYANIN-QUASI,AOBCS,Poghossian,ADiLOHBCS}.

More complicated boundary terms can be considered in the integrable theory
via a procedure demonstrated by Sklyanin\cite{SKLYANIN,CH}.
To begin with, let's impose  the following constrains on the $R$-matrix
\begin{align}
\mbox{Unitarity:~} &
  R_{12}(\lambda) R_{12}(-\lambda)= \zeta(\lambda)\id; \label{uni} \\
\mbox{Parity invariance:~} &
  P_{12}R_{12}(\lambda)P_{12}=R_{12}(\lambda);
\label{pari}\\
\mbox{Temporal invariance:~} &
  R_{12}(\lambda)^{t_{1}t_{2}}=R_{12}(\lambda);
\label{tempo} \\
\mbox{Crossing symmetry:~} &
  R_{12}^{t_{1}}(\lambda)R_{12}^{t_{1}}(-\lambda-2\rho)
  =\zeta(\lambda+\rho)\id;
\label{cross}
\end{align}
where $\zeta(\lambda)=\eta^2-\lambda^2$ and the crossing parameter $\rho=\eta$.  Here
$\id$ is the identity matrix, $P_{12}$ is the permutation operator,
$t_{\alpha}$ denotes transposition on the $\alpha$-th space.

According to the Sklyanin procedure, open boundary conditions
preserve the integrability of the model need to
provide 2-particle scattering in the bulk 'compatible' with the
scattering off the boundaries.  Such boundary conditions can be parametrized
by boundary matrices $K(\lambda)$ which satisfy the so called reflection
equations
\begin{equation}
\begin{aligned}
R_{12}(\lambda-\lambda') \stackrel{1}{K}_-(\lambda) R_{21}(\lambda+\lambda')
\stackrel{2}{K}_-(\lambda')
&= \stackrel{2}{K}_-(\lambda') R_{12}(\lambda+\lambda') \stackrel{1}{K}_-(\lambda)
R_{21}(\lambda-\lambda'),
\\
R_{21}(-\lambda+\lambda') \stackrel{1}{K_+^t}(\lambda) R_{12}(-\lambda-\lambda'-2\eta)
\stackrel{2}{K_+^t}(\lambda') &= \stackrel{2}{K_+^t}(\lambda') R_{21}(-\lambda-\lambda'-2\eta)
\stackrel{1}{K_+^t}(\lambda)  R_{12}(-\lambda+\lambda'),
\label{eqref}
\end{aligned}
\end{equation}
where $\stackrel{1}{K}(\lambda)=K(\lambda) \otimes \id$,
$\stackrel{2}{K}(\lambda)=\id \otimes K(\lambda)$, $K^t$ is the transpose
of $K$, and $R_{12}$ is a solution of the Yang-Baxter equation
(\ref{yangbaxter}).

General non-diagonal $\CC$-number solutions in case of the
rational $R$-matrix (\ref{rmatrix}) are the $K$-matrices~\cite{VeGo94}
\begin{equation}
\label{boundaries}
K_\pm(\lambda)=
\Matrix{cc}{\xi_\pm + \lambda &\lambda\mu_\pm\\
            \lambda\nu_\pm&\xi_\pm -\lambda}\; .
\end{equation}

Within this framework the analogue of the row-to-row transfer matrix
(\ref{transferT1}) as generator of commuting integrals of motion in the case
of open boundaries is the following operator \cite{SKLYANIN}
\begin{equation}
t(\lambda)=\tr_{\cal A}\left[ K_+(\lambda+\eta)\left[
T(-\lambda)\right]^{-1} K_-(\lambda)  T(\lambda)
 \right]\; .
\label{doubletransf}
\end{equation}
often called 'double row transfer matrix' in the literature.
For diagonal $K$-matrices the diagonalization of $t(\lambda)$ can be done by
an extension of the algebraic Bethe Ansatz \cite{SKLYANIN}.  For models based
on the rational $R$-matrices (\ref{rmatrix}) considered in this paper this
approach can be extended to certain non-diagonal $K$-matrices by
combination of suitable transformations in both quantum and auxiliary space
\cite{MRM,GM}.

\section{Quantum inverse scattering for single mode spin-boson model}
\label{spinboson}

In this section we apply the general theory presented in the previous section
to the spin-boson problem.  Starting from the $R$-matrix (\ref{rmatrix}) we
consider two types of Lax matrices satisfying (\ref{fundrel}):  the spin
degree of freedom is described by
\begin{equation}
\label{Spin-Lax}
{\cal L}_{j}^{(s)}(\lambda)=\left(
\begin{array}{cc}
       \lambda+\eta +\eta S_{j}^{z} & \eta S_{j}^{-} \\
       \eta S_{j}^{+} & \lambda+\eta - \eta S_{j}^{z}
       \end{array}\right), 
\end{equation} 
where the operators $
S_{j}^{z}$ and $S_{j}^{\pm}$ are chosen to be irreducible
representations of $su(2)$.  On way to construct a Lax matrix
for the bosonic degree of freedom is using a representations of
$su(2)$ in terms of Bose operators in (\ref{Spin-Lax}),
e.g. the Holstein-Primakoff or the Dyson-Maleev
representation. 
Here, however, we choose the bosonic Lax matrix as~\cite{KOREPIN-BOOK,BO} 
\begin{equation} 
\label{Boson-Lax}
       {\cal L}_{j}^{(b,1)}(\lambda)=\left( 
\begin{array}{cc}
       \lambda - \eta a_{j}^{\dag}a_{j}  & \beta a^{\dag}_{j} \\
       \gamma a_{j}  & -\frac{\beta\gamma}{\eta} \end{array}\right)\qquad ;\qquad
{\cal L}_{j}^{(b,2)}(\lambda)=
       \Matrix{cc}{\frac{\beta\gamma}{\eta} & \beta a^{\dag}_{j}\\
                  \gamma a_{j} &\eta a_{j}^{\dag}a_{j}+ \lambda +\eta}\ .\;
\end{equation}
with $[a_j,a_k^\dagger] = \delta_{jk}$.  For the construction of the transfer
matrix (\ref{doubletransf}) we also need the inverse of the bosonic
Lax operator which reads
\begin{equation}
\label{BosonInverse}
{\cal L}_{j}^{(b,1)^{-1}}(\lambda)={\det}_q[{\cal L}_{j}^{(b,1)}]^{-1}
              \Matrix{cc}{\frac{\beta\gamma}{\eta} & \beta a^{\dag}_{j}\\
                  \gamma a_{j} &\eta a_{j}^{\dag}a_{j}- \lambda +\eta}=
{\det}_q[{\cal L}_{j}^{(b,1)}]^{-1}{\cal L}_{j}^{(b,2)}(-\lambda)\ .
\end{equation}
Here $\det_q[{\cal L}_{j}^{(b,1)}]=-\beta\gamma({\lambda}/{\eta})$ is the
quantum determinant of ${\cal L}_{j}^{(b,1)}$.

The $R$-matrix (\ref{rmatrix}) intertwines both the spin and bosonic Lax matrices; additionally it is $GL(2)$ invariant, namely it commutes with
$\G\otimes\G$ for all $\G\in GL(2)$ (see (\ref{Gcondition})).
This $GL(2)$ invariance of the $R$-matrix implies
that $GL(2)$-transformed Lax matrices $G_1{\cal L} G_2$ are again solutions of the
Yang-Baxter relation (\ref{fundrel}).
At a formal level, $GL(2)$ transformations preserves the structure
of the spin Lax matrix (\ref{Spin-Lax}) and essentially
correspond to a rotation of the local spin.
We point out that this is not the case for the bosonic
Lax matrix (\ref{Boson-Lax});
though general $GL(2)$ transformations lead to canonical transformations
of the boson degrees of freedom, only
diagonal $GL(2)$ transformations preserve the structure of the Lax operator.

In what follows we restrict ourselves to setups with a single spin and
a single bosonic mode, corresponding to a two-site monodromy matrix.
Many relevant features are already contained in this simple case,
and an extension to the multi-spin and multi-mode situation is technically
straight forward.
Integrable spin-boson models can be constructed e.g. from the
transfer matrix for toroidal boundary conditions (periodic boundary
with two boundary twists)
\begin{equation}
t_{twist}(\lambda)=\trace_{\cal A}\left[ \G_s{\cal L}^{(s)}(\lambda-z_0)\G_b{\cal L}^{(b)}(\lambda-z_1) \right]
\end{equation}
where $\G_s,\G_b\in GL(2)$ do not depend on $\lambda$; the quantities 
$z_0,z_1$, shifting the spectral spectral parameter locally, are known as 
'impurities' in the jargon of QISM.   
Since $[t_{twist}(\lambda),t_{twist}(\mu)]=0$, in particular all 
coefficients of a $\lambda$-expansion of $t_{twist}(\lambda)$ commute with each other.
The first order coefficient of this $\lambda$-expansion of the
transfer matrix is a linear combination of the spin- and boson
operators occurring in the Lax matrix. This operator, where
spin- and boson degrees of freedom are decoupled, consequently commutes with
every Hamiltonian constructable from this transfer matrix.
This implies that
the problem can be effectively reduced to a {\em dressed} spin degree of
freedom, in which the Hamiltonian is block-diagonal~\cite{COHEN}.
This can be argued  from the $GL(2)$ symmetry of the $R$-matrix
discussed above, in that $\lambda$-independent twists are elements of
$GL(2)$ and can hence be absorbed in a redefined spin and boson Lax operator.
Note that this result is independent of the specific bosonic Lax
matrix chosen; either choice of a bosonic Lax operator, as given
in Eq.(\ref{Boson-Lax}) or obtained from
the spin Lax operator (\ref{Spin-Lax}) inserting e.g. the Holstein-Primakoff
or Dyson-Maleev bosonic representation of $su(2)$, will produce model
Hamiltonians with a conserved number operator $S^z+a^\dagger a$. 
Consequently all Hamiltonians
constructed in this way are of block diagonal form, with more or less
complicated non linearity in the bosonic and the interaction part. Examples
for such nonlinear generalizations of $H_{s-ph}$ can be found in
\cite{BUZEK,BO,JURCO,RYBIN,KUNDU}, where both twist matrices were chosen as
the same diagonal matrix.  Non-diagonal hermitian twist matrices have been
used in \cite{AH}.

For obtaining a model including counter rotating
terms from the XXX symmetric $R$ matrix, we look at open
boundary conditions.
This amounts to using the transfer matrix
(\ref{doubletransf}) as the generator of integrable Hamiltonians.
To this end we will consider the following monodromy matrix
\begin{equation}\label{mono-open}
\T(\lambda)=K_+(\lambda+\eta)
{\cal L}^{(s)^{-1}}(-\lambda-z_0){\cal L}^{(b)^{-1}}(-\lambda-z_1)K_-(\lambda)
{\cal L}^{(b)}(\lambda-z_1){\cal L}^{(s)}(\lambda-z_0)
\end{equation}
where the boundary matrices are taken from equation (\ref{boundaries}).

In the next section we discuss the exact solution of the eigenvalue problem for
the transfer matrix $t(\lambda)=\trace\T (\lambda )$ by means of the algebraic Bethe Ansatz
following two approaches.
First by 'direct' algebraic Bethe ansatz,
i.e. including the bosonic Lax operator (\ref{Boson-Lax}), and then
by first solving an auxiliary spin-spin problem and then performing a suitable
large-spin limit leading to the bosonic degree of freedom.
The explicit construction of model Hamiltonians from
(\ref{mono-open}) will be presented in Section \ref{hamiltonians}.

\section{Algebraic Bethe ansatz for the spin-boson transfer matrix}

The starting point of the exact diagonalization of the transfer matrix by
the algebraic Bethe ansatz is to identify a suitable eigenstate of the
transfer matrix, the so-called pseudo-vacuum $\ket{\Omega}$.  In the presence
of non-diagonal boundary matrices, this state can be identified by
a method proposed in Refs.~\onlinecite{MRM,RM}.
Sketching their approach,
the boundary matrices are brought to triangular form by a
suitable similarity transformation of the monodromy matrix (\ref{mono-open}).
This amounts to finding matrices $M_j=\Matrix{cc}{x_j&r_j\\y_j&s_j}$ in
\beqa
\label{gauge-t}
\tilde{\T}(\lambda)&=&M_0^{-1}\T(\lambda)M_0^{}\\
&=&M_0^{-1}K_+(\lambda+\eta)M_5^{}M_5^{-1}{\cal L}^{(s)^{-1}}(-\lambda-z_0)
M_4^{}M_4^{-1}{\cal L}^{(b)^{-1}}(-\lambda-z_1)M_3^{}M_3^{-1}K_-(\lambda)
M_2^{}\times\\
&&\ \times M_2^{-1}{\cal L}^{(b)}(\lambda-z_1)M_1^{}M_1^{-1}{\cal L}^{(s)}(\lambda-z_0)M_0
\eeqa
such that the lower left entry of both $M_0^{-1}K_+(\lambda+\eta)M_5^{}$ and
$M_3^{-1}K_-(\lambda)M_2^{}$ vanishes.
The remaining parameters in the $M_j$ have to be chosen such that the lower
left entries of the transformed Lax operators ${\cal L}^{(s,b)}$ annihilate the spin
and bosonic vacuum states $\ket{\omega_s}$ and $\ket{\omega_b}$ respectively.
As a result we can choose the product state $\ket{\Omega}= \ket{\omega_s}
\otimes\ket{\omega_b}$ as pseudo vacuum for the algebraic Bethe Ansatz.
Following Refs.~\onlinecite{MRM,RM} for the spin Lax matrix we readily obtain
${x_0}/{y_0}={x_1}/{y_1}$ and ${x_4}/{y_4}={x_5}/{y_5}$.
Note that having $M_j^{-1}{\cal L}^{(\alpha)} M_{j-1}
\ket{\omega_\alpha}$ triangular implies the same for $\tilde{M}_{j-1}
{\cal L}^{(\alpha)^{-1}}(\lambda-z_0) \tilde{M}_j \ket{\omega_\alpha}$, where the
matrices $\tilde{M}_j$ may differ from $M_j$ in their right column only.
Since the transfer matrix is independent of this
spurious freedom, we are free to choose $M_3=M_2$, $M_4=M_1$, and $M_5=M_0$
without loss of generality. If all Lax matrices are
spin Lax matrices, all $M_j$ can be chosen to be equal~\cite{MRM,RM},
corresponding to a global rotation of the spin.

\subsection{Algebraic Bethe ansatz with the bosonic Lax operator}
\label{ABA-direct}

In this section,
we adapt this procedure to the spin-boson model for the two-site
monodromy (\ref{mono-open}) and the choice
${\cal L}^{(b)}={\cal L}^{(b,1)}$;
the alternative choice ${\cal L}^{(b)}={\cal L}^{(b,2)}$
is discussed below. The approach described next for open boundary conditions
equally applies to toroidal boundary conditions.

The lower left entry of the
transformed bosonic Lax matrix
$\tilde{{\cal L}}^{(b)}(\lambda)\equiv M_2^{-1} {\cal L}^{(b)}(\lambda) M_1$
is (up to a rescaling of the spectral parameter)
\begin{equation}
\label{annihi-boson}
-x_1y_2\lambda + (\frac{\gamma}{\eta}x_2+y_2 a^\dagger)(\eta x_1 a-\beta y_1)
\end{equation}
Clearly, this operator can annihilate a vacuum state
for all $\lambda$ only if $x_1 y_2=0$.
The bosonic creation operator $a^\dagger$ has no right eigenstate, therefore
we have to choose $y_2=0$. This means that the gauge matrix left to
the bosonic Lax matrix must already be in upper triangular form.
Consequently, the pseudo-vacuum of the bosonic degree of freedom
is the coherent state $(\eta x_1a-\beta y_1)\ket{\omega_b}=0$ with
\begin{equation}
\frac{\beta}{\eta}\frac{x_0}{y_0}=\frac{\beta}{\eta}\frac{x_1}{y_1}
\equiv\frac{\beta}{\eta}\e^p\; .
\end{equation}
We also comment that for the Eq.(\ref{annihi-boson}) the  structure 
of two reference states found in\cite{YANG} is lost in the spin-boson case.

Writing  $\tilde{{\cal L}}^{(b)} = \Matrix{cc}{\tilde{A}^{(b)}&\tilde{B}^{(b)}\\
\tilde{C}^{(b)}&\tilde{D}^{(b)}}$ we have
\begin{equation}
\label{Lax-actionB}
\begin{aligned}
\tilde{C}^{(b)}\ket{\omega_b}&=0\,,\\
\tilde{A}^{(b)}\ket{\omega_b}&=\frac{y_1}{x_{2}}\left[(\lambda-z_1)\e^p\right]
\ket{\omega_b}\,, \\
\tilde{D}^{(b)}\ket{\omega_b}&= \frac{\beta\gamma}{\eta}
\frac{r_1\e^{-p}-s_1}{s_2}\ket{\omega_b}\,.
\end{aligned}
\end{equation}
The characteristic factor $\e^p$ is obtained from the condition that
$\tilde{K}_+=M_0^{-1}K_+\tilde{M}_0$ be upper triangular
\begin{equation}
\label{ratio}
\e^{-p}=\frac{K_{+;21}}{\kappa_+ -K_{+;22}}
\end{equation}
where $\kappa_+$ is an eigenvalue of $K_+$.  \\
Note that the parameters
$r_0,s_0,\tilde{r}_0,\tilde{s}_0$ as well as $r_2,s_2,\tilde{r}_2,\tilde{s}_2$
can be chosen such that both boundary matrices $K_+$ and $K_-$ are diagonal.
These parameters enter both the Bethe equations and the expression
for the eigenvalue ${\cal L}ambda(\lambda)$ of the transfer matrix,
Eq.~(\ref{Eigenvalue}).
The transfer matrix and its spectrum, nevertheless, is independent of
this choice.
This freedom provides with
a variety of isospectral Hamiltonians and corresponding Bethe equations.
The transfer matrix, however, differs from that obtained from
Eq.~(\ref{mono-open}) and corresponding diagonalized $K_+$ and $K_-$.
This is due to the fact that for  $r_j$, $s_j$ not being all the same,
the ``gauge'' transformations of the Lax operators do not induce
canonical transformations on the spin- and bosonic degrees of freedom.

Here, rather than bringing the boundary matrices into diagonal form, we choose
gauge transformations to simplify the Bethe ansatz equations.  To be specific,
let $\det M_j=x_j s_j-y_j r_j=1$ and $x_j=r_j=1$ for $j=0,1$ giving
$s_j=y_j+1=\e^{-p}+1$.
Furthermore, the condition $y_2=0$ for the existence of the
bosonic pseudo vacuum implies that the boundary matrix $K_-$ in
(\ref{gauge-t}) must be upper triangular (when bringing the monodromy matrix in
lower triangular form instead, the particular role of $x_j$ and $y_j$
is taken by $r_j$ and $s_j$ respectively).
Therefore, we are free to choose $M_2\equiv\id$ and the
action (\ref{Lax-actionB}) of
$\tilde{{\cal L}}^{(b)}(\lambda-z_1)$ on the bosonic coherent state simplifies
\begin{equation}
\label{Lax-actionb}
\begin{aligned}
\tilde{C}^{(b)}\ket{\omega_b}&=0\,,\\
\tilde{A}^{(b)}\ket{\omega_b}&=(\lambda-z_1)\ket{\omega_b}\,, \\
\tilde{D}^{(b)}\ket{\omega_b}&=-\ket{\omega_b}\,.
\end{aligned}
\end{equation}
Similarly, after the similarity transformation,
the action of the spin Lax operator ${\cal L}^{(s)}(\lambda-z_0)$ for a
$2S+1$ dimensional representation of $su(2)$ on $\ket{\omega_s}$ becomes
\begin{equation}
\label{Lax-actions}
\begin{aligned}
\tilde{C}^{(s)}\ket{\omega_s}&=0\,,\\
\tilde{A}^{(s)}\ket{\omega_s}&=(\lambda-z_0+\eta (S+1))\ket{\omega_s}\,, \\
\tilde{D}^{(s)}\ket{\omega_s}&=(\lambda-z_0-\eta (S-1))\ket{\omega_s}\,,
\end{aligned}
\end{equation}
while the boundary matrix $K_+$ is transformed into
\begin{eqnarray}
\tilde{K}_+&=&\e^{-p}\Matrix{cc}{-K_{+;21}\e^{2p}
+(2K_{+;11}-K_{+;22})\e^{p}+2K_{+;12} & *\\
0&K_{+;21}\e^{2p}-(K_{+;11}-2K_{+;22})\e^{p}-2K_{+;12}}\nonumber\\
&=&\Matrix{cc}{\kappa_+&*\\0&\kappa_+'}=:\Matrix{cc}{\xi_+ \pm \lambda\sqrt{1+\nu_+\mu_+}&*\\
                                 0&\xi_+ \mp \lambda\sqrt{1+\nu_+\mu_+}}\label{newK}
\end{eqnarray}
It is worth noticing that for a multi-site spin and multi-mode boson
generalization of this model the actions of all
consecutive Lax matrices on the pseudo-vacuum have to be in triangular form
for the above procedure to work.
Furthermore will we briefly discuss the diagonalization procedure,
when the bosonic
Lax matrix Eq.~(\ref{BosonInverse}) is inserted into the transfer matrix,
Eq.~(\ref{doubletransf}) -- and not Eq.~(\ref{Boson-Lax}) as above.
A straight forward calculation shows that then the gauge
matrix $M_1$ to the right of the bosonic Lax matrix and all following
Lax and boundary matrices must be in triangular form.
As a consequence, $K_+$ instead of $K_-$ must be triangular.

We now proceed to the result of the algebraic Bethe ansatz described above.
The eigenvalues $\Lambda(\lambda;\{\lambda_j\})$ of the transfer matrix
(\ref{doubletransf}) are obtained within the standard formalism~\cite{SKLYANIN}
\begin{equation}
\begin{aligned}
\Lambda(\lambda)&=-\left[\frac{(\lambda+\eta S+z_{0})(\lambda+\eta(S+1)-z_{0})
}{(\lambda+z_{0}+\eta S)(\lambda+z_{0}-\eta(S+1))}
\right]
\left[\frac{\lambda-z_{1}}{\lambda +z_{1}}
\right]
 \\
&\times \left[\frac{2(\lambda+\eta)(\lambda+
\zeta_{-})(\lambda+\zeta_{+})}{(2\lambda+\eta)} \right]
\prod_{j=1}^{n}
\frac{(\lambda-\lambda_{j}-\frac{\eta}{2})}{(\lambda-\lambda_{j}+\frac{\eta}{2})}
\frac{(\lambda+\lambda_{j}-\frac{\eta}{2})}{(\lambda+\lambda_{j}+\frac{\eta}{2})}
 \\
& +\left[\frac{(\lambda-\eta S+z_{0})(\lambda-\eta(S-1)-z_{0})}{(\lambda+z_{0}+\eta S)(\lambda+z_{0}-\eta(S+1))}
\right]
\left[\frac{\lambda+z_{1}+\eta }{\lambda +z_{1}} \right]
 \\
& \times \left[\frac{2\lambda(\lambda-
\zeta_{-}+\eta)(\lambda-\zeta_{+}+\eta)}{(2\lambda+\eta)} \right]
\prod_{j=1}^{n}
\frac{(\lambda-\lambda_{j}+\frac{3\eta}{2})}{(\lambda-\lambda_{j}+\frac{\eta}{2})}
\frac{(\lambda+\lambda_{j}+\frac{3\eta}{2})}{(\lambda+\lambda_{j}+\frac{\eta}{2})},
\label{Eigenvalue}
\end{aligned}
\end{equation}
The pseudo-vacuum $\ket{\Omega}$ has eigenvalue $\Lambda_0(\lambda) =
\Lambda(\lambda;\emptyset)$. The consequence of choosing ${\cal L}^{(b,2)}$
as bosonic Lax operator consists in an overall minus sign and
$z_1\rightarrow -z_1$.
Eigenstates are characterized by the roots
$\{\lambda_j\}$ of the Bethe equations
\begin{equation}
\begin{aligned}
-\frac{(\lambda_i+\eta (S-\frac{1}{2})+z_0)(\lambda_i+\eta
(S+\frac{1}{2})-z_0)}{(\lambda_i-\eta
(S+\frac{1}{2})+z_0)(\lambda_i-\eta (S-\frac{1}{2})-z_0)}
&\left[\frac{\lambda_i-z_1-\frac{\eta}{2}}{\lambda_i+z_1+\frac{\eta}{2}}\right]
\frac{(\lambda_i+ \zeta_+ - \frac{\eta}{2})(\lambda_i+\xi_- -
\frac{\eta}{2})}
{(\lambda_i-\zeta_+ + \frac{\eta}{2})(\lambda_i-\xi_- + \frac{\eta}{2})}\\
=\prod_{j\neq i}
\frac{(\lambda_i-\lambda_j+\eta)
  (\lambda_i+\lambda_j+\eta)}{(\lambda_i-\lambda_j-\eta)
  (\lambda_i+\lambda_j-\eta)}\;.
\label{BetheEquations}
\end{aligned}
\end{equation}
where the boundary matrix $K_+$ has been rescaled with $\sqrt{1+\mu_+
\nu_+}$ and $\zeta_+:=\xi_+/\sqrt{1+\mu_+ \nu_+}$ is the inverse of
the effective boundary field.  Eqs.~(\ref{BetheEquations}) coincide with the
Bethe equations for an open chain with diagonal boundary matrices after a
rotation of the spin and bosonic degrees of freedom.
Choosing ${\cal L}^{(b,2)}$ as the bosonic Lax operator induces the change
$z_1\rightarrow -z_1$.

We close this paragraph commenting on the general implications
of the restriction to triangular $K_-$.
Though the boundary conditions do break the $su(2)$ symmetry of the system,
there is a remaining $u(1)$ conserved charge in the system in this case.
Hence, the spin boson models which can be diagonalized within this
approach should not be expected to feature counter rotating terms.
Nevertheless for non-diagonal triangular $K_-$, the resulting model
will show some features of counter rotating terms.
We emphasize, however, that the triangularity of $K_-$ is a
constraint necessary \emph{only} for this method for the diagonalization of
the transfer matrix which is not related to the integrability of the
corresponding spin-boson model.

\subsection{Solution by algebraic contraction of the spin-spin transfer matrix}
\label{ABA-contraction}

It is well known that bosonic degrees of freedom can be realized as
spin operators, in a certain limit. Such a limit constitutes an example of algebraic contraction\cite{GIL}. In this section we demonstrate how
this idea can be applied to the different stages of QISM method, thus
allowing to obtain integrable spin-boson model from certain auxiliary
integrable
spin model. The  approach is pursued here in the case
the integrable model is with open boundary conditions, thus extending
the theory for the toroidal boundaries\cite{AH}.

At the level of the Lax matrices, the spin to boson mapping is achieved by the combination of
two singular transformations, one in the auxiliary and the other
in quantum space, to have a finite result in the limit.
Explicitly, the bosonic Lax matrix is obtained as~\cite{AH}
\begin{equation}
{\cal L}^{(b)}(\lambda)=\lim_{\epsilon \rightarrow \infty}
k(\epsilon) {\cal L}^{(J)}(\lambda-\delta),
\label{limit}
\end{equation}
 where the matrix  $k(\epsilon)$ is a diagonal matrix
\begin{equation}
\label{triks}
k(\epsilon)=\diag\left(1,\frac{1}{\epsilon^2} \right)\;,
\delta=\eta+ \eta\epsilon^{2}/2 ,  J= \epsilon^{2}/2\;,
\end{equation}
and the quantum space transformation is defined by $\left\{
J^{-},\frac{1}{\epsilon^2} J^{+}, J^{z} \right\}
\rightarrow \left\{ a^{\dag},a, -a^{\dag} a +\frac{\epsilon^{2}}{2} \right\}$.
We have assumed
$\beta=\gamma=\eta$ in the definition of the Bose Lax matrix,
Eq.~(\ref{Boson-Lax}) for convenience.

We apply the Eq.\ref{limit} to an auxiliary double-row transfer
matrix $t_{a}(\lambda)$ that in the limit of $\epsilon \rightarrow \infty $
is equal to $t(\lambda)$
\begin{eqnarray}
t_{a}(\lambda)= \tr \left \{ K_+(\lambda+\eta) \left[{\cal L}^{(s)}(-\lambda-z_{0}) \right]^{-1} \left[ {\cal L}^{(J)}(-\lambda-\delta-z_{1})
\right]^{-1} \left[ k(\epsilon) \right]^{-1} K_-(\lambda) \right .
\nonumber \\
\left .\times k(\epsilon) {\cal L}^{(J)}(\lambda-\delta-z_{1}) {\cal L}^{(s)}(\lambda-z_{0})  \right \}.
\end{eqnarray}

The above operator $t_{a}(\lambda)$ can be conveniently
rotated into $\bar{t}_{a}(\lambda)$ through a new transformation $U_j$
(\ref{quantumtransf}) acting in the quantum spaces, along the line 
suggested by the property (\ref{invariance}). This simplifies the  
the effect of $k(\epsilon)$, acting on the auxiliary space. 
The transformed 
$\bar{t}_{a}(\lambda)$ reads
\begin{eqnarray}
\bar{t}_{a}(\lambda)&=&U t_{a}(\lambda)U^{-1} \nonumber \\
&=&\tr \left \{ k(\epsilon) K_+(\lambda+\eta) \left(k(\epsilon)\right)^{-1} \left[{\cal L}^{(s)}(-\lambda-z_{0}) \right]^{-1}
\left[ {\cal L}^{(J)}(-\lambda-\delta-z_{1}) \right]^{-1} K_-(\lambda) \right. \nonumber  \\
&\times& \left.  {\cal L}^{(J)}(\lambda-\delta-z_{1})
{\cal L}^{(s)}(\lambda-z_{0})  \right \} \label{topenN}
\end{eqnarray}
Therefore, we can obtain  the spectral properties of 
$t_{a}(\lambda)$ by the analysis of  $\bar{t}_{a}(\lambda)$, whose
eigenvalues were previously obtained by means of algebraic Bethe
ansatz \cite{MRM}. In this rephrasing the $K$-matrix we need to take into 
account is 
\begin{eqnarray}
\widetilde{K}_{+}(\lambda)&=&k(\epsilon)K_{+}(\lambda)(k(\epsilon))^{-1} \nonumber \\
 &=& \left(
\begin{array}{cc}
        \xi_{+}+\lambda &  \epsilon^2 \lambda \mu_{+} \\
        \frac{1}{\epsilon^2} \lambda\nu_{+} & \xi_{+}-\lambda
        \end{array}\right).
\end{eqnarray}

Of course, the eigenvalues of the transfer matrix $t_{a}(\lambda)$ and $\bar{t}_{a}(\lambda)$ are identical and  given by the following expression
\begin{eqnarray}
\begin{aligned}
\Lambda_{a}(\lambda) &=\left[\frac{(\lambda+\eta S+z_{0})(\lambda+\eta+\eta S-z_{0})
}{(\eta S+\lambda+z_{0})(\eta(S+1)-\lambda-z_{0})}
\right]
\left[\frac{(\lambda-\delta+\eta J+ \eta -z_{1}) (\lambda+\delta+\eta J+z_{1})}{(\eta J +\lambda+\delta+z_{1})
(\eta(J+1)-\lambda-\delta-z_{1})} \right]
 \\
&\times\left[\frac{2(\lambda+\eta)(\lambda+ \zeta_{-})(\lambda+
\zeta_{+})}{(2\lambda+\eta)} \right] \prod_{j=1}^{n}
\frac{(\lambda-\lambda_{j}-\frac{\eta}{2})}{(\lambda-\lambda_{j}+\frac{\eta}{2})}
\frac{(\lambda+\lambda_{j}-\frac{\eta}{2})}{(\lambda+\lambda_{j}+\frac{\eta}{2})}-
 \\
&-\left[\frac{(\lambda-\eta S+z_{0})(\lambda+\eta-\eta S-z_{0})}{(\eta S+\lambda+z_{0})(\eta(S+1)-\lambda-z_{0})}
\right]
\left[\frac{(\lambda-\delta-\eta J+\eta -z_{1}) (\lambda+\delta-\eta J+z_{1})}{(\eta J +\lambda+\delta+z_{1})
(\eta(J+1)-\lambda-\delta-z_{1})} \right]
 \\
& \times \left[\frac{2\lambda(\lambda-
\zeta_{-}+\eta)(\lambda-\zeta_{+}+\eta)}{(2\lambda+\eta)} \right]
\prod_{j=1}^{n}
\frac{(\lambda-\lambda_{j}+\frac{3\eta}{2})}{(\lambda-\lambda_{j}+\frac{\eta}{2})}
\frac{(\lambda+\lambda_{j}+\frac{3\eta}{2})}{(\lambda+\lambda_{j}+\frac{\eta}{2})},
\end{aligned}
\end{eqnarray}
while the Bethe ansatz equations are given by
\begin{eqnarray}
\label{Bethe-spin}
&&-\left[\frac{(\lambda_{j}-\frac{\eta}{2}+\eta S+z_{0})(\lambda_{j}+\frac{\eta}{2}+\eta S-z_{0})}{(\lambda_{j}-\frac{\eta}{2}-\eta S+z_{0})(\lambda_{j}+\frac{\eta}{2}-\eta S-z_{0})}
\right] \left[\frac{(\lambda_{j}+\delta-\frac{\eta}{2}+\eta
J+z_{1})(\lambda_{j}-\delta+\frac{\eta}{2}+\eta J-z_{1})}{(\lambda_{j}+\delta-\frac{\eta}{2}-\eta
J+z_{1})(\lambda_{j}-\delta+\frac{\eta}{2}-\eta J-z_{1})} \right] = \\
&& \left(\frac{\lambda_{j}- \zeta_{-}+\frac{\eta}{2}}{\lambda_{j}+
\zeta_{-}-\frac{\eta}{2}} \right)
\left(\frac{\lambda_{j}-\zeta_{+}+\frac{\eta}{2}}{\lambda_{j}
+\zeta_{+}-\frac{\eta}{2}} \right)\prod_{\stackrel{i=1}{i\neq
j}}^{n}
\frac{(\lambda_{j}-\lambda_{i}+\eta)}{(\lambda_{j}-\lambda_{i}-\eta)}
\frac{(\lambda_{j}+\lambda_{i}+\eta)}{(\lambda_{j}+\lambda_{i}-\eta)},
\nonumber
\end{eqnarray}
where $\displaystyle \zeta_{\pm}=-\frac{\xi_{\pm}}{\sqrt{1+\mu_{\pm}\nu_{\pm}}}$, which do not depend on $\epsilon$. For the algebraic Bethe ansatz,
one of the two following assumptions on the parameters
$\xi_{\pm}, \mu_{\pm}$ and $\nu_{\pm}$ has been made (see \cite{MRM})
\begin{eqnarray}
\frac{1+\sqrt{1+\nu_{-}\mu_{-}}}{\mu_{-}}&=&\frac{1+\sqrt{1+\nu_{+}\mu_{+}}}{\mu_{+}\epsilon^2}, \mbox{  or} \label{const1} \\
\frac{-\nu_{-}}{1+\sqrt{1-\nu_{-}\mu_{-}}} &=&\frac{1+\sqrt{1+\nu_{+}\mu_{+}}}{\mu_{+}\epsilon^2} \label{const2}
\end{eqnarray}

To obtain the eigenvalues and the Bethe ansatz equation of the spin-boson problem, associated to the
transfer matrix $t(\lambda)$ we take the limit
$\epsilon \rightarrow \infty$ of Eqs.(\ref{Bethe-spin}).
This leads to the same eigenvalue (\ref{Eigenvalue})
and Bethe ansatz equations (\ref{BetheEquations}) as obtained from the
'direct' ansatz.
In the limit of $\epsilon\to\infty$ the constraints
(\ref{const1}),(\ref{const2})
become $\nu_{-}=0$ and $\nu_{-} \mu_{-}=0$, respectively.
This means that the boundary matrix $K_{-}(\lambda)$ has to be in
a triangular (or even diagonal) form already, and
$\zeta_-=-\xi_-$.

Considering $\mu_{\pm}=\nu_{\pm}=0$ in the eigenvalue expression
and Bethe ansatz equation above we can recover the results
obtained for the case of diagonal $K_{\pm}(\lambda)$ $K$-matrices
without use of the limit process.

As final remark we observe that a definite limit $\epsilon\rightarrow 0$
exists, despite the matrices $k(\epsilon)$ are singular in the limit;
such singularity is compensated by the divergence in the ``impurity'' $\delta$
(see Eq.(\ref{triks})).

\section{Integrable Hamiltonians}
\label{hamiltonians}

The transfer matrix $t(\lambda)$ for rational solutions of the YBE
is a polynomial expression
in the spectral parameter. Because the transfer matrix is a commuting family
of operators in $\lambda$, the coefficients
of such a polynomial are the integrals of the motion of the theory.
In the following section we exploit this property
of the transfer matrix to extract the Hamiltonian.
In the section \ref{quasi-classical} we obtain the Hamiltonian following an alternative procedure that is in the spirit of
the quasi-classical expansion.

\subsection{Integrable models derived from the transfer matrix}
\label{direct}

We analyze the integrals of the motion generated by the
transfer matrix (\ref{doubletransf}) and the bosonic Lax matrix
${\cal L}^{(b)}={\cal L}^{(b,2)}$ for spin-1/2.
We therefore expand it in powers of the spectral parameter
\begin{equation}\label{def-Hams}
t(\lambda)=:\sum_{n=0}^6 H_n\lambda^n
\end{equation}
where we set $\eta=1$.

There are only two independent constants of the motion in this problem
and we find that $H_6$, $H_5$, and $H_0$ are constants and the relations
\begin{eqnarray}\label{Hams}
2 H_4-H_3&=&5\mu_-\nu_+\\
H_4-H_2+H_1&=&3\mu_-\nu_+
\end{eqnarray}
With this result for the general case we are now ready for a systematic
case study.

At first, we focus on diagonal boundary matrices $K_\pm$.
In this case we find up to a constant
\begin{equation}
H_4=4\beta\gamma(S^z-\hat{n})\; .
\end{equation}
This constant of the motion readily unveils this Hamiltonian to be of
the Jaynes-Cummings type in the RWA  because the Hilbert space splits up in
two-dimensional (spin-1/2) subspaces with fixed eigenvalue of $H_4$.
Problems in absence of a conserved number operator in general will require
more sophisticated techniques as employed for the diagonalization of
the XYZ chain or the XXZ chain with non-diagonal
boundary fields~\cite{CAO,NEPO}.

For diagonal $K_+$ and
non-diagonal but triangular $K_-$,
the expansion (\ref{def-Hams}) leads to (up to a constant)
\begin{equation}
\label{Htriag}
H_4=\Omega_0(S^z-\hat{n})
+ C_1 \left(1-\frac{1}{2s}+2S^zs\right)S^+ +C_2 \hat{a} +C_3\hat{a}(2S^z-\hat{n})
\end{equation}
where $s$ is the spin, $\Omega_0=4\beta\gamma$,
$C_1=\nu_-(2\xi_+ +2z_0-1)$, $C_2=2\gamma\nu_-(z_1+\xi_+)$,
and $C_3=2\gamma\nu_-$.
As we will show below, the Hilbert space of the
non-hermitian Hamiltonian (\ref{Htriag}) is not reduced to two-dimensional
subspaces; therefore, it describes a resonant  spin-boson model (i.e with vanishing de-tuning of  the spin-boson frequencies)
containing counter-rotating terms. 
It can be treated by the method used above.
A careful analysis reveals that the
non-hermiticity of the Hamiltonian is inherent to this transfer matrix unless
the spin degrees of freedom are decoupled from the boson.  We attribute this
feature to the structure of the bosonic Lax matrix (\ref{Boson-Lax}) used here.

In what follows we restrict ourselves to spin $s=1/2$.
A vacuum state for the Bethe ansatz on the Hamiltonian (\ref{Htriag})
then is $\ket{0,\frac{1}{2}}$, using the notation $\ket{n,\sigma}$
such that $S^z\ket{n,\sigma}=\sigma\ket{n,\sigma}$ and
$\hat{n} \ket{n,\sigma}=n\ket{n,\sigma}$.
Defining $C_2=:C_1+\delta$, $C_3=C_1+\delta+\Delta$,
the creation operator $B(\lambda)$ has the form
\beqa
B(\lambda)&=&\left[\Delta^2\lambda(\lambda+1)-\delta^2+2\delta\Delta S^z\right]
\left\{\hat{a}^\dagger(\hat{n}-\xi_-)+\hat{n}(\hat{n}-1)\right\}\nonumber\\
&&-\Delta\left[2\delta\hat{n}+2\Delta(\Delta\lambda(\lambda+1)-\delta)\hat{a}
+2\Delta\delta\hat{a}\hat{n})\right]S^-\nonumber\\
&& -\Delta(\Delta(\lambda^2+\lambda+\xi_-)-\delta\xi_-) S^-
\eeqa
A right-eigenstate
$\ket{\lambda_1,\dots,\lambda_N}:=
B(\lambda_1)\dots B(\lambda_N)\ket{0,\frac{1}{2}}$ involves
all states $\ket{n,\sigma}$ with $n\leq N$ except $\ket{N,-\frac{1}{2}}$.
It is interesting to notice that for positive integer values of the
inverse boundary field $\xi_-$ the Hilbert space accessible
from the vacuum $\ket{0,\frac{1}{2}}$ is restricted to
\beq\label{integerxi}
{\cal H}_{\xi_-\in\NN^+}:=\{\ket{n,\sigma}\; ;\; n\leq \xi_-\}
\eeq

Without loss of generality we set $g_1=1$ and
$\mu_+=\nu_+=0$.
The latter corresponds to a diagonal boundary matrix
$K_+(\lambda)=\xi_+\id+\lambda\sigma_z$;
any non-diagonal $K_+$ can be diagonalized by a gauge
transformation\cite{MRM,RM}
and proper $r_0,s_0,\tilde{r}_0,\tilde{s}_0$.
For convenience, we choose $\gamma=\beta=1$ and rename $g_2=:g$.
The most general Hamiltonian generated in this way is equivalent to
\beq\label{genHam}
H=-\frac{g_1 H_4+g_2 H_1}{4g_1}=:H_s+H_b+H_{sb}+H_{sbb}
\eeq
via suitable canonical transformations on the spin and bosonic degrees
of freedom.

\begin{figure}[ht]
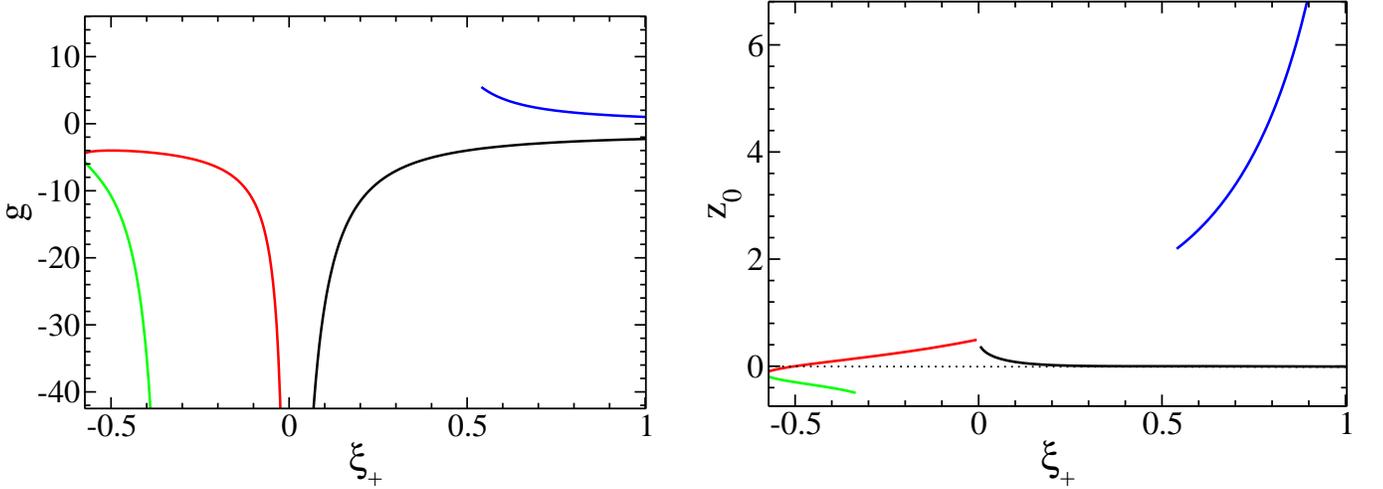

\begin{center}
\includegraphics[angle=0,width=0.48\textwidth]{ImplicitPlot.eps}
\hfill
\includegraphics[angle=0,width=0.48\textwidth]{z0Plot.eps}
\caption{{\em Left}: Solutions of the nonlinear equation in $\xi_+$ and $g$
emerging from inserting equation (\ref{HermBoson}) into (\ref{HermSpin}).
{\em Right}: the corresponding values for $z_0$.
\label{ImplicitFigure}
}
\end{center}
\end{figure}

Our goal will be to have the spin and boson Hamiltonian together with major
part of the spin-boson interaction hermitian.  To this end, we set
$\nu_-=\frac{\mu_-}{\xi_+}(\frac{1}{g}+\xi_+z_1)$, which makes the
occurring terms $\hat{a}S^z$ and $\hat{a}^\dagger S^z$ hermitian and the choice
$\xi_-=1-z_1-\frac{2z_0 - 2\xi_+ -1}{2z_0 + 2\xi_+ -1}$ realizes the same for
$\hat{a}S^-$ and $\hat{a}^\dagger S^+$.  $z_1=1/2$ eliminates a term
$\hat{n}S^+$ and the pure spin Hamiltonian becomes hermitian if
\begin{equation}\label{HermSpin}
g=4-\frac{4}{\xi_+}+16\frac{2\xi_+-1}{2z_0-6\xi_+ -1}\; .
\end{equation}
The terms $\hat{a}$ and $\hat{a}^\dagger$ in the pure boson part of the
Hamiltonian are obtained hermitian if
\begin{equation}\label{HermBoson}
z_0=\frac{1}{2}\pm\frac{\sqrt{1-g(g+2)\xi_+^2+g^2(g+1)\xi_+^4}}
{\sqrt{g}(g\xi_+^2-1)}\;
\end{equation}
and it turns out that both conditions (\ref{HermSpin}) and (\ref{HermBoson})
are compatible and fixes a one-dimensional manifold in $\xi_+$--$z_0$ space
(see figure \ref{ImplicitFigure}).
\vspace*{5mm}

The resulting Hamiltonian is as follows
\begin{eqnarray}\label{simpleHam}
H_s &=&\left(4\frac{1-4\xi_+^2}{\xi_+}-4\frac{1+2\xi_+}{2z_0+2\xi_+-1}
-8\frac{\xi_+(2\xi_+ -1)}{2z_0 -6\xi_+ -1}\right)\; S^z
+2\mu_-\frac{(\xi_+ -1)((2z_0-1)^2-4\xi_+^2)}{\xi_+(2z_0 -6\xi_+ +1)}\; S^x \\
H_b &=& g^2\frac{4\xi_+^2}{g\xi_+^2 -1}\; \hat{n}
-g\frac{\mu_-\xi_+ (g\xi_+ +2)}{g\xi_+^2 -1}\;  (\hat{a}^{}+\hat{a}^\dagger)
+g^2\frac{\mu_-\xi_+^2}{g\xi_+^2 -1}\;  \hat{a}\hat{n}\\
H_{sb}&=&-8g\xi_+\;  \hat{n}S^z + 2\mu_-(g\xi_+ +2)\;
(\hat{a}^{}+\hat{a}^\dagger)S^z +2g(2z_0-2\xi_+ -1)\;
(\hat{a}^{}S^- +\hat{a}^{\dagger}S^+)\\
H_{sbb}&=&g\mu_-\left((2z_0+2\xi_+ -1)\frac{g\xi_+ +2}{2g\xi_+}\;
\hat{a}^{\dagger^2}S^+ -(2z_0-2\xi_+ -1)\; \hat{a}^2 S^-\right)\nonumber\\
&&\quad +g(2z_0+2\xi_+ -1)
(\mu_- \;\hat{n}^2 S^+ + 2\;\hat{a}^\dagger \hat{n} S^+)
-4g\mu_-\xi_+\;\hat{a}\hat{n} S^z
\end{eqnarray}
Besides the hermitian parts $H_s$ and $H_{sb}$, there are many non-hermitian
terms, in particular nonlinear couplings as in $H_{sbb}$.
It is worth focusing on the non-hermitian term in $H_{b}$ proportional
to $\hat{a}\hat{n}$. This term has an immediate physical interpretation:
it describes a leakage of bosons/photons from the cavity, where the
escape rate is proportional to the number of bosons/photons
(i.e. the intensity of the cavity boson field).

\subsection{Integrable models derived from the quasi-classical limit
of the transfer matrix}
\label{quasi-classical}

Another way to obtain an integrable Hamiltonian is to take
the so called quasi-classical limit of the
transfer matrix~\cite{Sklyanin,Kulish,HIKAMI}.
It consists in a series expansion in the 'quantum parameter' $\eta$
of the transfer matrix around $\eta=0$:
$
\hat{t}(\lambda)= \hat{\tau}^{(0)} +
\eta\,\hat{\tau}^{(1)}(\lambda) + \eta^2 \,\hat{\tau}^{(2)} (\lambda) +\dots
$
with the aim of creating a commuting family of quasi-classical
transfer matrices $\tau^{(k)}(\lambda)$.
This procedure may be particularly useful for extracting 'simple'
though non-locally interacting Hamiltonians out of the transfer matrix.
Examples are the Gaudin magnets and corresponding BCS-like models.

There is a wide freedom of introducing an $\eta$-dependence
to the boundary matrix parameters in the theory,
without destroying its integrability.
For the case of integrable spin chains with diagonal
boundaries\cite{HIKAMI} this procedure was a valid tool to define
integrable one-parameter extension of Gaudin
models in non uniform magnetic fields\cite{ADiLOHBCS};
such models describe metallic grains with pairing and
magnetic interaction\cite{AOBCS}.

In the present case the transfer matrix is the finite sum
\begin{equation}
\hat{t}(\lambda)= \eta^{-k} \hat{\tau}^{(-k)}+ \dots + \hat{\tau}^{(0)} +
\eta\,\hat{\tau}^{(1)}(\lambda) + \eta^2 \,\hat{\tau}^{(2)} (\lambda) +\dots +\eta^m \,\hat{\tau}^{(m)} (\lambda)
\label{quasi-transfer}
\end{equation}
where $k$ and $m$ are integers.
Expanding the commutator relation $[\hat{t}(\lambda),\hat{t}(\lambda')]=0$ in $\eta$,
we obtain
\[[\hat{t}(\lambda),\hat{t}(\lambda')]=\sum_{l=-2k}^{2m} \eta^l C_l(\lambda,\lambda') = 0\;,\]
which implies $C_l(\lambda,\lambda')=0$ for all $l$.
The first relevant terms are
\begin{align}\label{commexp}
C_{-2k}(\lambda,\lambda')=&[\hat{\tau}^{(-k)}(\lambda),\hat{\tau}^{(-k)}(\lambda')]\;,\nonumber\\
C_{-2k+1}(\lambda,\lambda')=&[\hat{\tau}^{(-k)}(\lambda),\hat{\tau}^{(-k+1)}(\lambda')] +
[\hat{\tau}^{(-k+1)}(\lambda),\hat{\tau}^{(-k)}(\lambda')]\;, \nonumber \\
C_{-2k+2}(\lambda,\lambda')=&[\hat{\tau}^{(-k)}(\lambda),\hat{\tau}^{(-k+2)}(\lambda')]
+[\hat{\tau}^{(-k+2)}(\lambda),\hat{\tau}^{(-k)}(\lambda')]
+[\hat{\tau}^{(-k+1)}(\lambda),\hat{\tau}^{(-k+1)}(\lambda')]\;,\nonumber \\
\end{align}
>From the expressions above, one finds that the first
$\hat{\tau}^{(n)}(\lambda)$ which is not a $\CC$-number (times the identity)
gives rise to a family of commuting operators.
Generically, the lowest order $\hat{\tau}^{(-k)}$ is a $\CC$-number. Therefore
the first class of integrable models were generated
by $[\hat{\tau}^{(-k+1)}(\lambda),\hat{\tau}^{(-k+1)}(\lambda')]=0$.
In the presence of boundary matrices, these are typically non-trivial operators
but representing non-interacting Hamiltonians.
The task is then to tune the free parameters such that $\hat{\tau}^{(-k+1)}(\lambda)$
is also a $\CC$-number, and that the lowest non-trivial order in $\eta$,
e.g. $\hat{\tau}^{(-k+2)}(\lambda)$, be an interesting Hamiltonian.

Here we
allow $\xi_\pm$, $\mu_\pm$, $\nu_\pm$ to be generic functions of
$\eta$ whose form can be fixed to modify to our convenience the
$\eta$-expansion of the transfer matrix, and ultimately the Hamiltonian.
Specifically, we write the boundary parameters as
\begin{equation}
x_\pm=x_\pm^{(-1)}\eta^{-1}+ x_\pm^{(0)}+ x_\pm^{(1)}\eta
\end{equation}
with $x=\{\xi,\mu,\kappa\}$, and
set $\mu_\pm^{(-1)}=\nu_\pm^{(-1)}=\xi_+^{(-1)}=0$. Then
the lowest non-vanishing term in the expansion, proportional to $\eta^{-2}$,
already contains non trivial operators;
they can be made vanishing by the choice $\mu_+^{(0)}=\nu_-^{(0)}=0$.
Analogously,
the next term in the expansion can be made proportional to the
identity by $\nu_+^{(0)}=\xi_+^{(0)}=0$ and $z_1=0$.
The next non-trivial term in the $\eta$-expansion $\hat{\tau}^{(0)}(\lambda )$
can be used as Hamiltonian
\begin{eqnarray}
H  \doteq \frac{1}{\lambda^2 \alpha \beta} \hat{\tau}^{(0)}(\lambda ) &=& 4 \Omega_0a^\dagger a +
2\alpha \left (a+a^\dagger \right )+ \Delta_z S^z+
4  \gamma z_0 \alpha S_x \nonumber \\
&&+ G S^z a^\dagger a -4\gamma z_0 (a S^-+a^\dagger S^+) +
\gamma^2 \left [S^+ S^-+S^- S^+ -2 (S^z)^2\right ]+C \id
\end{eqnarray}
where $\Omega_0=-\left (\lambda^2-z_0^2\right )$, $\alpha=\beta \nu_-^{(1)}-\gamma \nu_+^{(1)}$ and
$\Delta_z=-4 \lambda^2 -2\gamma^2 (1-2\xi_+^{(1)})$, $G = 4\gamma^2(1-\gamma)$ and
$C=-\gamma z_1 \left \{\Omega_0  \left [2 (2 +\xi_-^0 -\alpha \beta \nu_-^1) \lambda^2  +2 \xi_-^0 \lambda +
 \right ]-2 (\lambda-z_0)(\gamma-\lambda^2 \xi_-^0 )\right \}$.

For $S=1/2$ the Hamiltonian simplifies to
\begin{equation}
H_{1/2}=\Omega_0 a^\dagger a+ \tilde{\Delta}_z
 S^z +G S^z a^\dagger a+G \alpha S^z \left (a+a^\dagger \right )-  g
\left ( S^+ a^\dagger + S^- a\right )
\label{quasi-Hamiltonian}
\end{equation}
where the bosonic operators have been displaced $a\rightarrow a+\alpha/2$,
$\tilde{\Delta}_z=\Delta_z + \alpha^2 G$ and a term proportional to the identity
was omitted.
The Hamiltonian (\ref{quasi-Hamiltonian}) display the rotating and counter-rotating simultaneously. This can be evidenced by a rotation $R$ of the spin
\begin{eqnarray}
R H_{1/2} R^{-1}=\Omega_0 a^\dagger a+\tilde{\Delta}_z \left [\cos(2\theta)S^z-\sin(2\theta) S^x\right ]+ G \left [\cos(2\theta)S^z-\sin(2\theta) S^x\right ]a^\dagger a  \nonumber \\
-{g}\frac{\cos(2\theta) +1}{2}
\left ( S^+ a^\dagger + S^- a\right ) -{g}\frac{\cos(2\theta) -1}{2}
\left ( S^+ a + S^- a^\dagger \right ) \;.
\end{eqnarray}
where $R\doteq \exp [\theta (S^+ -S^-)]$ and
$\tan \theta=G\alpha/g$ has been chosen to eliminate the $S^z (a+a^\dagger)$ term.

\section{Conclusions}
\label{conclusion}

By the Quantum Inverse scattering method we have constructed
integrable models with twisted and open boundary conditions which contain
rotating as well as counter rotating terms.

Twisted boundary conditions
lead to Hamiltonians where solely rotating ({\it or} counter-rotating) terms appear.

In case of open boundaries, we  diagonalize the models by
algebraic Bethe ansatz when at least one of the boundary matrices
has triangular form.
To this end, we extended the procedure
for spin chains to models which are derived from a particular bosonic
Lax matrix besides the standard spin Lax matrix.
We further analyzed generic boundary matrices
in the sense that they cannot be brought both in triangular form
at the same time.
For this choice of boundary fields,
the symmetry leading to conservation of  the 'number operator' $S^z+\sum_j
a^\dagger_j a_j$ is broken.

Taking the full transfer matrix the  integrable Hamiltonians
are non-hermitian.
However, a parameter choice is found that shifts the non-hermiticity to
non-linear terms in the spin-boson interaction and the bosonic part of
the Hamiltonian. In the latter, the non-hermiticity can be interpreted
as a density dependent leakage of photons out of the cavity.
Different parameter choices could be analyzed in order
that non-hermitian terms will appear elsewhere, e.g. in the spin part
of the Hamiltonian.
It is worth studying what type of non-hermiticity are most realistic
in order to tailor the parameters closest to this situation.
The diagonalization of the resulting Hamiltonians together with
a discussion of the effects of the undesired non-hermitian parts is left to
forthcoming work.

Interestingly enough, by a procedure that is close in spirit to the  
quasi-classical limit, also hermitian Hamiltonians can be extracted out 
of the transfer matrix.
We have given an explicit example,
which contains both rotating and counter-rotating terms.
The evidence we have for this is that any eigenstate (i.e. coherent states)
of the bosonic part is spread out all over the bosonic Hilbert space
by means of the interaction term.

Though integrable counter-rotating spin-boson models have been formulated
in this work, it must be emphasized that their diagonalization with e.g.
the algebraic Bethe ansatz is still an open problem.
It could be interesting to study general open boundary conditions still for
the XXX model but based on the Holstein-Primakoff representation of $su(2)$.
The advantage of such an approach is that the Hamiltonian is constructed
in a straight forward manner from the existing solution of the open
$XXX$ spin chain Hamiltonian and hence also the resulting
spin-boson Hamiltonian is hermitian by construction.
The challenge again lies in the diagonalization of the transfer matrix.
Finally, the analysis of spin-boson models fanning out from a bulk with $XYZ$
symmetry (parametrized by elliptic functions) constitute an interesting 
challenge for future investigation.

\section*{Acknowledgments}
We thank A. Kundu for fruitful discussions;
G.A.P. Ribeiro thanks FAPESP (Funda\c c\~ao de Amparo \`a
Pesquisa do Estado de S\~ao Paulo)
for financial support and the DMFCI for the hospitality.

\end{document}